\documentclass[journal, onecolumn]{IEEEtran}

\usepackage{algorithmic} 
\usepackage{algorithm}
\usepackage{amsmath}
\usepackage{ifpdf}
\usepackage{graphicx}
\usepackage{graphics}
\usepackage{subfigure}
\usepackage{balance}
\usepackage{cite}
\usepackage{cases}

\begin{document}
	
	\title{Towards Efficient Dynamic Virtual Network Embedding Strategy for Cloud IoT Networks}
	
	\author{\IEEEauthorblockN{Duc-Lam Nguyen, HyungHo Byun, Naeon Kim, Chong-Kwon Kim}
	\IEEEauthorblockA{ \\
		Dept. of Computer Science and Engineering \\
		Seoul National University, Seoul, Republic of Korea \\}}
	\maketitle
	
	\begin{abstract}
	Network Virtualization is one of the most promising technologies for future networking and considered as a critical IT resource that connects distributed, virtualized Cloud Computing services and different components such as storage, servers and application. Network Virtualization allows multiple virtual networks to coexist on same shared physical infrastructure simultaneously. One of the crucial keys in Network Virtualization is Virtual Network Embedding, which provides a method to allocate physical substrate resources to virtual network requests. In this paper, we investigate Virtual Network Embedding strategies and related issues for resource allocation of an Internet Provider(InP) to efficiently embed virtual networks that are requested by Virtual Network Operators(VNOs) who share the same infrastructure provided by the InP. In order to achieve that goal, we design a heuristic Virtual Network Embedding algorithm that simultaneously embeds virtual nodes and virtual links of each virtual network request onto physic infrastructure. Through extensive simulations, we demonstrate that our proposed scheme improves significantly the performance of Virtual Network Embedding by enhancing the long-term average revenue as well as acceptance ratio and resource utilization of virtual network requests compared to prior algorithms.
		
	\end{abstract}
	
	\begin{IEEEkeywords}
		Cloud Computing, Virtual Network Embedding, Network Virtualization, Resource Allocation
	\end{IEEEkeywords}

\section{Introduction}
Nowadays, Cloud Computing\cite{1} is a large scale distributed model for providing on demand IT resources such as storage, servers, networks and application as a service through data centers or high-speed networks. More and more Cloud Computing centers are facilitated with virtualization technologies to make better use of server's capacity and to guarantee higher flexibility and reliability. The concept of Network Virtualization has provided a promise for the current and future Infrastructure as a Service(IaaS) provision in cloud networks. Network Virtualization (NV) \cite{2,3} has become a popular concept in a wide range of technologies. NV enables network operators to share their physical infrastructure with different tenants by forming multiple virtual networks. In the cloud environment of Network Virtualization, the Internet Service Provider(ISP) is separated into two different roles, namely the Infrastructure Providers (InPs) and Service Providers (SPs). Infrastructure Providers build, maintain and manage the physical infrastructure while Service Provides are owners of virtual networks. Service Providers deploy different protocols as well as offer multiple end-to-end different services to the end user of the cloud services\cite{4}.

\begin{figure}[t]
	\centering
	\includegraphics[width=0.5\linewidth]{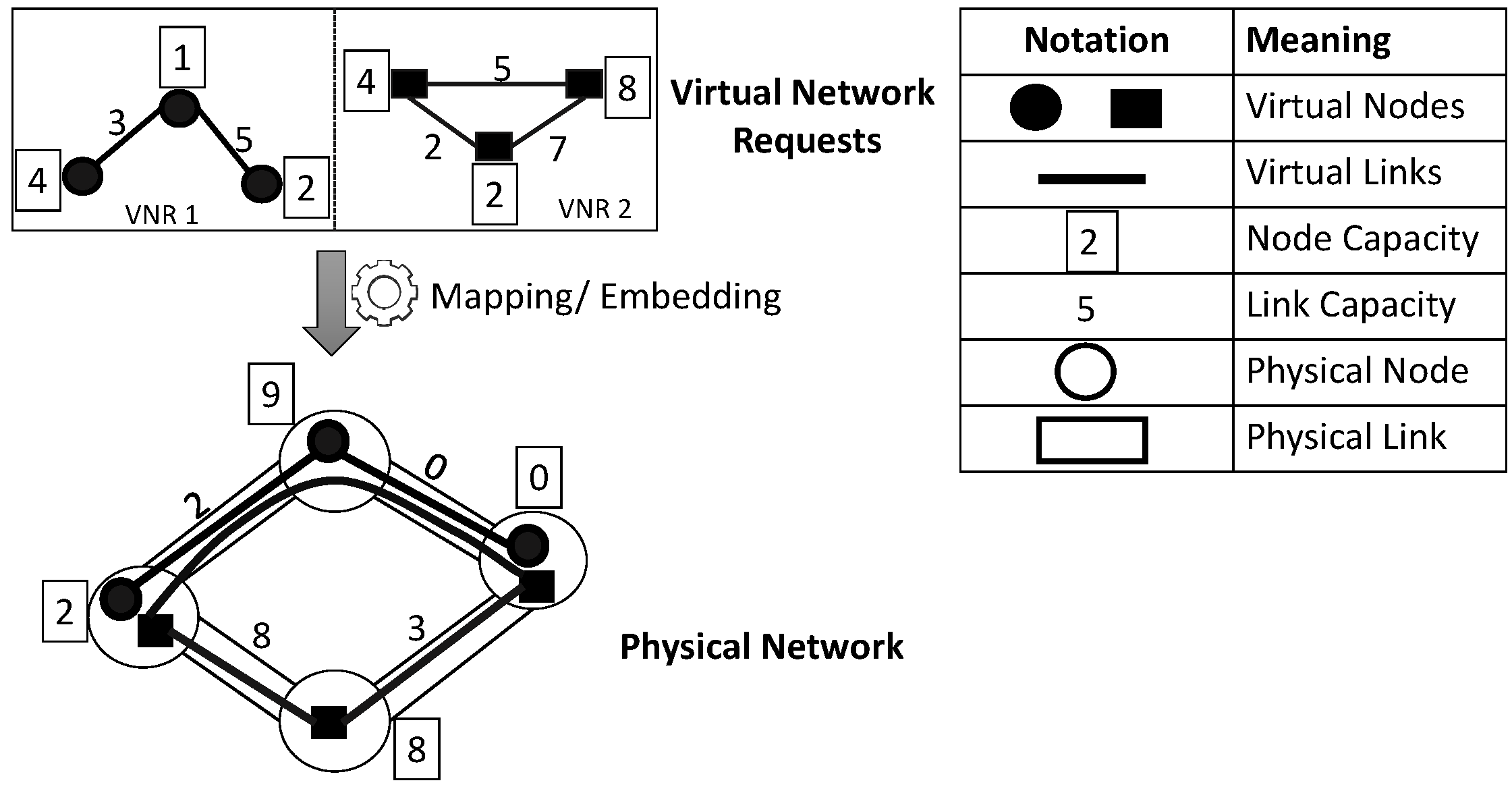}
	\caption{Virtual Network Embedding problem}
	\label{fig:figure1}
\end{figure}

Network Virtualization brings many potential benefits. First, it offers a virtual environment to design and evaluate new network architectures or protocols. Second, it enables resource sharing among virtual networks and optimizes resource allocation of physical infrastructure. In Network Virtualization, a virtual network (VN) is a combination of virtual nodes and virtual links on top of a substrate network (SN). A virtual node can be hosted on any available substrate node and interconnected by a set of virtual links. Virtual links can be spanned over several substrate links, forming a virtual topology. Service Providers send virtual network requests with the information of virtual nodes and virtual links capacity requirements to InPs who map the virtual network requests onto substrate network in turn, as shown in Figure. \ref{fig:figure1}. In cloud infrastructures, virtual networks are offered as a service to interconnect distributed cloud components and resources to support vary Cloud-based services for customers and users. As a result, the cloud service request can be abstracted as a Virtual Network Request(VNR) where virtual nodes stand for virtual machines and virtual links stand for cloud infrastructure links. And VN provision can be achieved via cloud systems such as Cloud Stack, OpenStack using Software Defined Networking controllers and underlying data plan system OpenVSwitch.

A major challenge of Network Virtualization technology is known as Virtual Network Embedding (VNE) problem  that deals with efficient allocation of substrate resources to virtual requests. Therefore, an efficient utilization of substrate network resources absolutely depends on the VNE algorithms under constraints such as node and link resources. The node constraints consist with several factors such as CPU capacity, memory and storage capacity or even the location of nodes in the substrate topology. The link constraints can be bandwidth capacity, delay, throughput, jitter and so on. Consequently, the VNE can be divided into two stages: virtual node mapping stage which assigns the virtual nodes to the substrate nodes and virtual link mapping stage where the virtual links are mapped onto physical links connecting these corresponding nodes after the virtual node mapping stage. Solving the VNE problem is NP-hard \cite{5,6,7} and it can be reduced to multi-way separator problem\cite{8}.

Recently, the VNE problem has received attention significantly to design diversified Internet architectures. The primary goal of VNE researches is to maximize the economic benefit (called long-term average revenue) by accommodating as many as Virtual Network Requests (VNRs) onto the same shared physical network\cite{9,10,11,12}. However, the revenue of InPs depends on the VNRs and in real-world situations because VNRs are dynamic and unpredictable. It is known as the problem with online VNRs. However, most prior works have just focused on static VNRs. Also, most of previous researches separated the VNE problem into two independent phases. That is, the virtual nodes are mapped onto substrate nodes at the first phase according to the local resource of substrate nodes without caring the other attributes such as location of substrate nodes or available bandwidth between nodes while these parameters have significant influence on the VNE algorithms' performance. Then, the virtual links are mapped simply using the shortest path algorithm\cite{9}. However, the separation of virtual node mapping and virtual link mapping might lead neighboring virtual nodes to be separated widely in the physical topology. Besides, this separation increases the cost of virtual link mapping phase. So, the acceptance ratio, as well as long-term revenue might be decreased.

\begin{figure*}[t]
	\centering
	\includegraphics[width=0.5\linewidth]{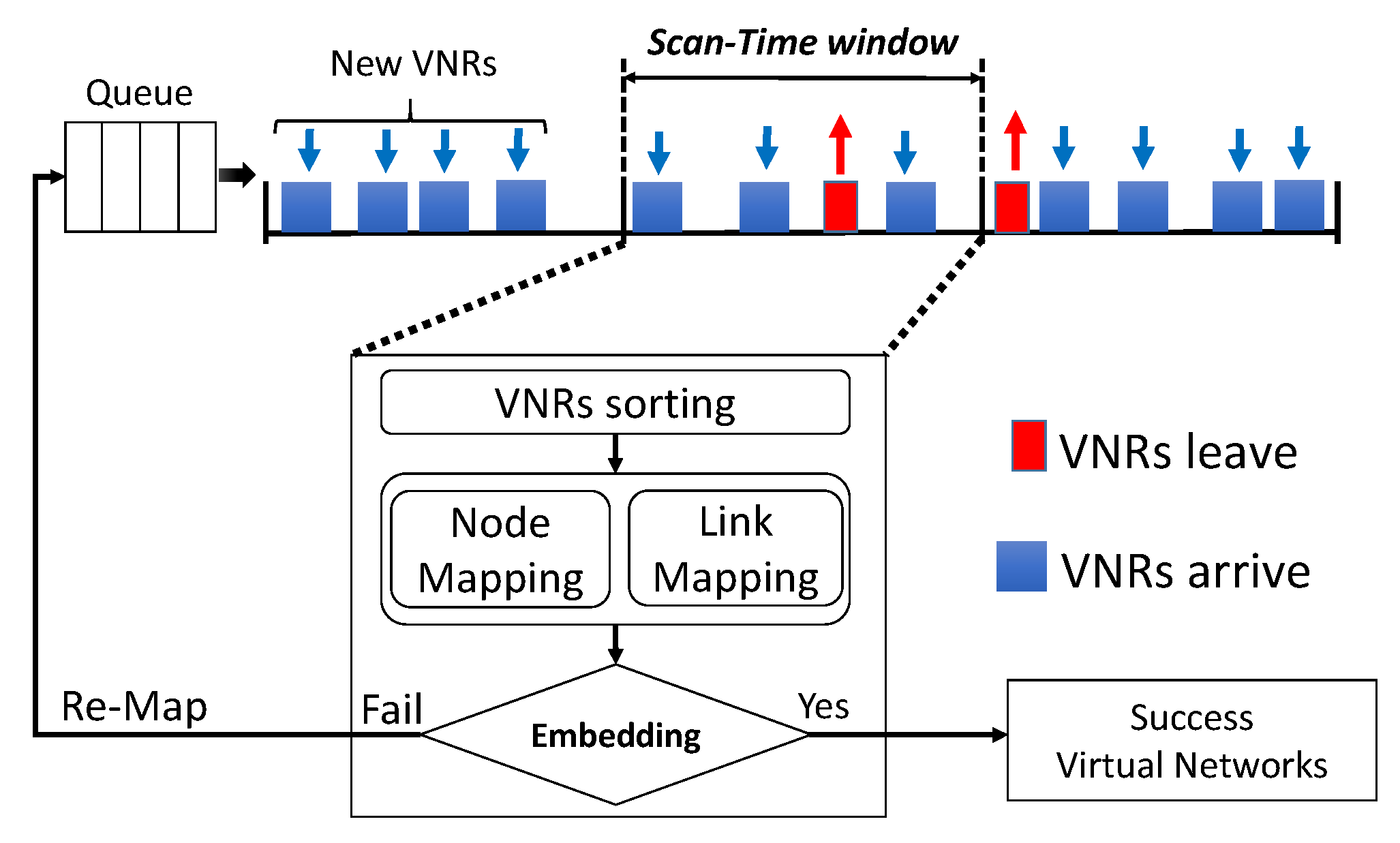}
	\caption{The overview of RT-VNE algorithm}
	\label{fig:figure2}
\end{figure*}

With the aforementioned motivations, in this paper, we investigate a new strategy to coordinate virtual link mapping and virtual node mapping in terms of handling online VNRs. The overview of our strategy is shown in Figure. \ref{fig:figure2}, the arriving VNRs are handled within scan-time intervals, if there is any VNR which can not be processed within a scan-time window, it is deferred to try in next scan-time windows. In order to maximize the long-term revenue of InPs, we propose a new strategy to handle VNRs in each scan-time window, the VNRs are grouped, analyzed and ordered following CPU capacity and bandwidth capacity demand. And, we also consider the resources of substrate network simultaneously to optimize resource allocation. The benefit of proposed algorithm lies in that it can further leverage the dynamic VNRs and thus help benefit the revenue of InPs. The key idea of our algorithm is that our proposed algorithm first choose X embedding candidates where X is the search number, and in each candidate embedding, we choose a substrate node as the parent node (root node) which is mapped to the first VN node which is chosen by a selection rule. Then, VN nodes and links are mapped simultaneously based on joint node and link constraint.  Finally, we choose the one has the best quality and feasible for embedding. The detail of proposed algorithm is described in later section.
Through extensive simulations, we demonstrate that our proposed algorithm outperforms the prior algorithms\cite{6,9,11}.  

The main contributions of this paper can be summarized as follows: 
\begin{enumerate}
	\item We find out the limitation of existing schemes. Without ability to adapt dynamic VNRs and the separation in virtual node and link mapping phases, the existing schemes may perform inefficiently in case of dynamic and large scale networks.
	\item A new VNE algorithm is proposed. We introduce a new real-time Virtual Network Embedding strategy which coordinates virtual node and virtual link mapping while handling dynamic virtual network requests as in real-world cloud environment. In which, the key point lies on the method we use to handle group of VNRs in each scan-time window.
	\item Through extensive experiments, we show that the proposed scheme achieves a significant improvement compared to prior algorithms in terms of acceptance rate, resource utilization and long-term average revenue.  
\end{enumerate}

The rest of paper is organized as follows: related works and issues are discussed in Section II. In section III, we present the VNE problem formulation. In section IV, we present our real-time Virtual Network embedding algorithm. Then, we evaluate our proposed VNE algorithm in section V. Finally, Section VI concludes overall of this paper.

\section{Related works}

The VNE problem has received attention significantly in academic or industrial area and presented the main resource allocation challenge in virtual cloud infrastructure. There are many studies to find out efficient strategies to maximize the benefit for InPs including subgraph isomorphism\cite{12}, path splitting and migration\cite{9}, topology-aware node ranking\cite{13} or other heuristic, meta-heuristic algorithms\cite{14,15,16} and so on. Generally, the existing studies can be clarified into two primary categories: online approaches or offline approaches.

The offline algorithms process based on the assumption that all virtual network requests are known and defined in advance. Beck et al.\cite{17,18} proposed a distributed and parallel VNE framework named DPVNE, which can be used in combination with various cost-reducing embedding algorithms. Through evaluation results, they proved that DPVNE achieves smaller message overhead and suggests a trade-off between message overhead and parallelism level. Fajjari et al.\cite{19} introduced a Max-Min Ant Colony Meta-heuristic algorithm which uses parallel artificial ants to iteratively explore the search space for feasible virtual network request candidates and select the candidates based on several specific resource constraints. The work of J.Lu et al.\cite{20} considered the single VNE in specific backbone-star topologies and assumed the substrate resources are unlimited with only bandwidth constraints. In these offline VNE approaches, the VNR can be considered to gather for embedding, which may achieve a better performance compared to online approaches. However, offline VNE approaches do not contemplate the possibility of re-mapping existing VNRs to enhance the performance of the embedding the substrate network. The arrival of VNR is unpredictable in real environment, where dynamic algorithms are more suitable and preferred. 

The dynamic VNE approaches study the on-demand VNR assignment problem where algorithms dynamically compute a feasible set of substrate nodes and links to embed the virtual nodes and links upon the arrival of VNRs. Chowdhury et al.\cite{11} proposed a coordinate node and link mapping strategy by formulating the VNE as a MIP, in which the node mapping is achieved by facilitating the link mapping in the subsequent phase. Randomized and deterministic rounding techniques are used to guarantee better correlation between node and link mapping phases .Yu. Minlan et al.\cite{9} proposed the rethinking design of the substrate network to enable simpler mapping algorithms and more efficient use of physical resources, without restricting the problem space. They first designed the substrate network to split a virtual link to span over multiple substrate paths. And then, they employed path migration to periodically re-optimize the utilization of the substrate network to support  the accommodating of new requests. In this research, a virtual link can be mapped to multiple substrate paths for splittable flows and path migration is used to optimized periodically the utilization of physical resources. X.Cheng et al.\cite{20} handled VNE problem by using the PageRank algorithm, where both substrate nodes and virtual nodes are ranked based on metrics and then the virtual network is embedded based on these ranks. The goal is to increase the acceptance ratio of online virtual network requests as well as overall revenue. 

On the other hand, the authors\cite{5} provided a heuristic algorithm based on a centralized algorithm which tends to maintain a low and balanced of both substrate nodes and substrate links. A distributed virtual network mapping algorithm was proposed by Houdi et al.\cite{22}. In this research, the VNRs can be decomposed in hub-and-spoke clusters and these clusters can be embedded independently, so it achieves to reduce the complexity of VNE. However, this approach has lower performance as well as efficiently compared to centralized approaches.

Besides, in cloud environments, the VNE problems also subject to dynamic changes and have to deal with node and link failures\cite{23,24,25}. Houidi et al.\cite{23} and Gou et al.\cite{24} proposed adaptive and and survivable embedding algorithms to deal with link failures where the objective is to minimize the embedding cost. There are several types of failures such as physical node failure, physical link failure and virtual link failure and so on.
\begin{table}[t]
	\small\sf\centering
	\caption{Terminology used through this paper.\label{T1}}
	\begin{tabular}{p{2cm} p{5cm}}
		\hline
		\textbf{Notation} & \textbf{Description}  \\ 
		\hline
		$G^S$ & Substrate Network \\
		$N^S$ & Set of Substrate Nodes \\
		$L^S$ & Set of Substrate Links \\
		$G^V$ & Virtual Network Requests \\
		$N^V$ & Set of Virtual Nodes \\
		$L^V$ & Set of Virtual Links \\
		$R^N$ & Physical Node resources \\
		$R^L$ & Physical Link resources \\
		$C_{Req} (n^V)$ & CPU request of VN node $n \in N^V$ \\
		$C_{RemA} (n^S)$ & CPU available remaining of SN Node $n^S \in N^S$ \\
		$Cap^{VN}(n^V)$ & VN-Node-Capacity \\
		$Cap^{SN}(n^S)$ & SN-Node-Capacity \\
		$Ltr(l^S)$ & Link-traffic-ratio \\
		$BW_{Req}^V (l^V)$ & BW requested for VN link $l^V \in L^V$ \\
		$BW_{RemA}^S (l^S)$ & Available BW of SN link $l^S \in L^S$  \\	
		\hline
	\end{tabular}
\end{table}			 

\section{System Modeling}
\label{sec:systemmodeling}
In this section, we present general embedding problem. The notations used in this context are summarized in Table \ref{T1}. 			
\subsection{Network Models}
\quad \textbf{Substrate Network:} A physical can be described as a weighted undirected graph $G^S = (N^S,L^S)$, where $N^S$ is the set of substrate of nodes and $L^s$ is the set of substrate links. Each substrate link $l^S(n^S_i, n^S_j) \subset L^s$ interconnects two substrate nodes $n^S_i$ and $n^S_j$. We denote the set of loop-free substrate paths by $P^S. $The $C^S(n^S)$ and $BW^S(l^S)$ refer to total available CPU capacity of substrate node $n^S$ and total available bandwidth of substrate link $l^S$, respectively. An example of substrate network is shown in lower part of Figure. \ref{fig:figure1}, where the numbers in square are the attribute of nodes indicated by the available CPU capacity. The numbers beside the edges are the attribute of links indicated by the available bandwidth capacity.  	

\textbf{Virtual Network:} Similar to the substrate network, we denote a Virtual Network Request by $G^V=(N^V,L^V)$, in which $ N^V$ is set of virtual nodes and $L^V$ is set of virtual links. $C^V_{Req}(n^V)$ and $BW^V_{Req}(l^V)$ are CPU capacity and bandwidth capacity requirement of virtual nodes and virtual links, respectively. For example, as VNRs are shown in upper part Figure. \ref{fig:figure1}, where the numbers in square are the attribute of virtual nodes indicated the CPU capacity demands and the numbers beside the edges are the attribute of virtual links indicated the bandwidth demands.
\subsection{Virtual Network Embedding problems}
The VNE problem is defined by an embedding $\Gamma$  from a virtual network $G^V$ to a subset of the substrate network $G^S$. The embedding can be expressed as follows: 
\begin{equation}
\Gamma : G^V \mapsto (N^{S*}, P^{S*}, R^N, R^L)
\end{equation}
where $N^{S*} \subset N^S$, $P^{S*} \subset P^S$, and $R^N$ and $R^L$ are physical node and link resources allocated for VNRs.	
The VNE problem can be divided into two categories: 1) virtual node mapping, 2) virtual link mapping.	
Firstly, each virtual node of a VNR is mapped to substrate network as follows: 
\begin{equation}
1)   \hspace{10mm}  \Gamma^N : N^V \mapsto N^{S*} 
\end{equation}
we have $\Gamma^N(n^V) \in N^{S*}$ with $n^V \subset N^V$ subject to the CPU capacity constraint as follows:	
\begin{equation}
	\begin{aligned}
	C_{Req}(n^V) \leq& \ C^S(\Gamma^N(n^V)) \\
	C_{RemA}(n^S) =& \ C(n^S) - \sum_{n^V \subset N^V} {C_{Req}(n^V)}
	\end{aligned}
\end{equation}
where $C_{Req}(n^V)$ is the amount of CPU capacity requested by the virtual node $n^V$; $C_{RemA}(n^S)$ is the remaining CPU capacity available  of substrate node $n^S$ after embedding virtual node $n^V$. Secondly, each virtual link can be mapped to paths connecting the corresponding nodes of substrate network. We denote virtual link mapping as follows: 
\begin{equation}	
2) \hspace{10mm}	\Gamma^L : L^V \mapsto P^{S*}
\end{equation}
where $\forall{l^V(n^V_i, n^V_j) \subset L^V}$, and $P^S(\Gamma^N(n^V_i)$ is set of substrate path connecting virtual node $n^V_i$ and $n^V_j$ $(\Gamma^N(n^V_j) \subseteq \Gamma^L(l^V))$ subject to bandwidth capacity constraints: 
\begin{equation}
	\begin{aligned}
	BW^V_{Req}(l^V) \leq& \ BW^S(p), \forall p \in \Gamma^L(l^V)  \\				
	BW^S(p) =& \ \min_{l^S \subset p} BW^S(l^S)  \\				
	BW^S_{RemA}(l^S) =& \ BW^S(l^S) - \sum_{l^V \subset L^V} BW^V_{Req}(l^V)
	\end{aligned}
\end{equation}
where $BW^V_{Req}(l^V)$ is requested bandwidth capacity for virtual link $l^V \subset L^V$. $BW^S_{RemA}(l^S)$ denotes the amount of available bandwidth capacity remaining of substrate link $l^S$. 

\subsection{Performance Metrics}

In this paper, the following metrics are used to evaluate the performance of our proposed algorithm. First, we define the $\textit{long-term average revenue}$ $R(G^V)$ where the Internet Provider obtains from the Service Provider for serving a VNR $g^V \subset G^V$, as follows:
\begin{equation}
maximize \quad \lim_{T \to \infty}\frac{\int_{t_a}^{t_e} \sum_{i=1}^{k} R_{g^V}^{i}(t)dt}{T}
\end{equation}
where $k$ is the number of virtual network requests which are accepted successfully by substrate network in duration of the VNR being served in the substrate network. And $t_d = t_e-t_a$ with $t_a$ to denote the arriving time of a VNR, $t_e$ is expiration time of the VNR. In this article, we aim to maximize the long-term average revenue. 
\begin{equation}
R^i_{g^V} = \left[\sum_{n^V \in N^V} C^V(n^V) + \varPsi \sum_{l^V \in L^V} BW(l^V)\right]*t_d
\end{equation}

Second, in each time window, the ratio of the number of successful embedded virtual network to all VNRs is defined as \textit{Acceptance Ratio}. The variable $x \in \{0,1\}$, if the $i^th$ VNR is accepted successfully, $x_i=1$, otherwise $x_i=0$.  
\begin{equation}
{AcptRatio = \left[\sum_{i \in VNR^{accept}} x_i\right] / \sum |VNRs| * 100\%}
\end{equation}

Third, higher substrate resource utilization has always been an important objective. Resource utilization includes three different types: node utilization, link utilization and network resource utilization. 

\begin{equation}
	\begin{aligned}
	NodeU& = \dfrac {\sum_{n^V \in N^V_{m}} C(n^V) } { \sum_{n^S \in N^S} C(n^S)} \\ 
	LinkU& = \dfrac{\sum_{l^V \in L^V_{m}} BW(l^V)}{\sum_{l^S\in L^S} BW(l^S)} \\
	NetU& = \dfrac{\sum_{  n^V \in  N^V_{m}} C(n^V) + \sum_{ l^V \in  L^V_{m}} BW(l^V)} {\sum_{ n^S \in  N^S} C(n^S) + \sum_{l^S\in L^S} BW(l^S)}
	\end{aligned}
\end{equation}

Besides, there are also unpopular metrics such as total allocated bandwidth, delay and throughput, the number of active physical nodes or links and so on. 
\section{Virtual Network Embedding Algorithm}
\label{sec:VNE}
\begin{figure*}[t!]
	\centering
	\frame{\includegraphics[width=0.75\linewidth]{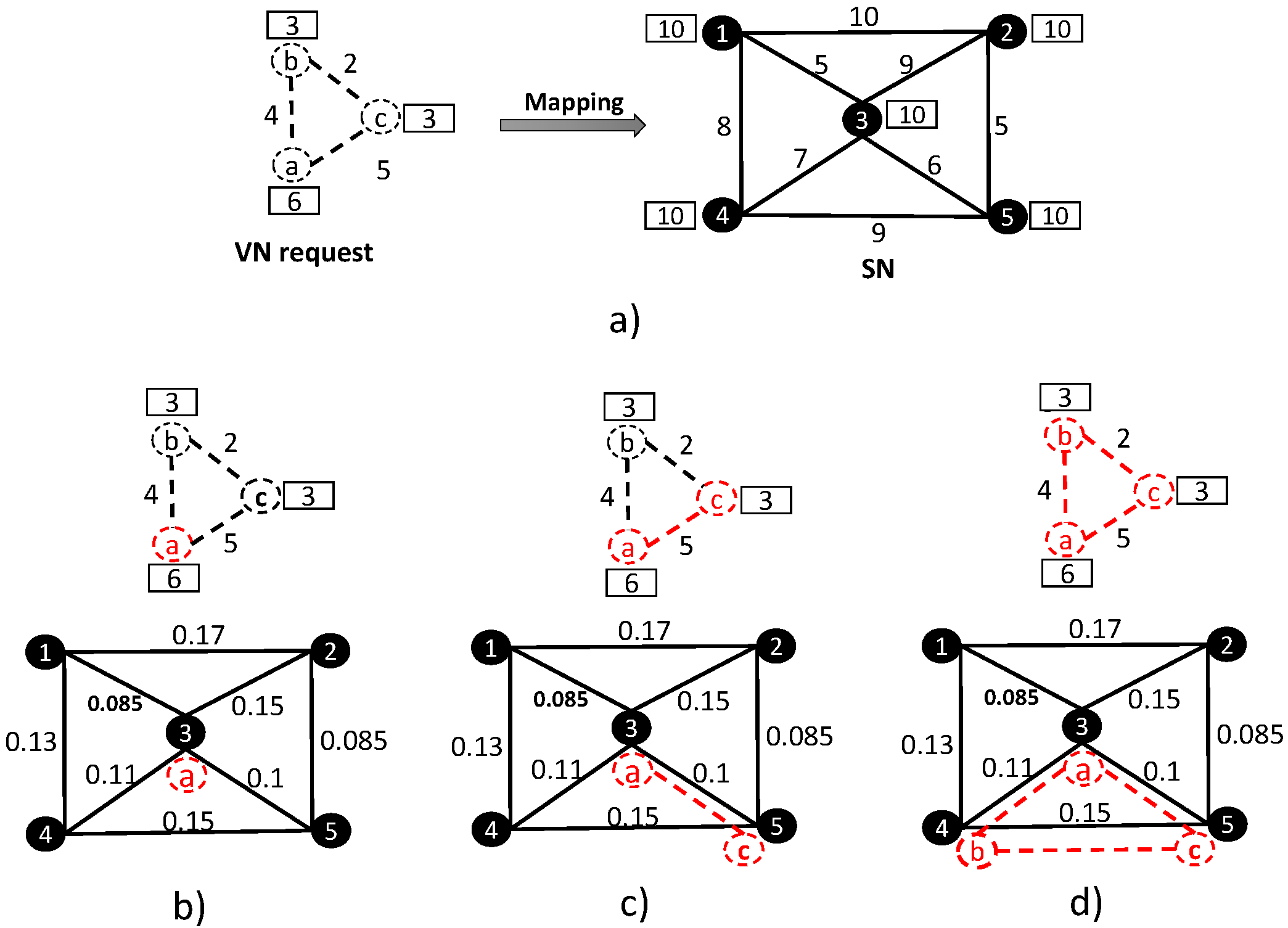}}
	\caption{RT-VNE algorithm Example}
	\label{fig:figure3}
\end{figure*}

In this section, we introduce our real-time VNE strategy. The overview of our proposed algorithm is shown in Figure. \ref{fig:figure2}, and the conjunction of virtual node-link mapping is depicted clearly.

\subsection{Online Virtual Network Request Handling}
In reality, the VNRs dynamically arrive and stay in substrate network an interval time before leaving. In order to handle online VNRs, we divide time into scan-time windows (interval times) which consider how long the VNRs is executed frequently (i.e., an hour or several days). In detail, VNRs in each scan-time window are processed sequentially (Figure. \ref{fig:figure2}).

\begin{algorithm}[t!]
	\caption{: RT-VNE Algorithm}
	\begin{algorithmic}[1]
		\STATE \textbf{Input:}  $G^V, G^S$ \\
		\STATE \textbf{Output}: Resource Efficient Mapping Solution
		\STATE \textbf{Step 1:} RT-VNE sorts virtual nodes in each VNR according to (Eq. 10) and select parent virtual node for each VNR.
		\STATE \textbf{Step 2:} Similarly, RT-VNE selects parent substrate node to map with selected parent virtual node (\textbf{Step 1}) according to (Eq. 11).
		\STATE \textbf{Step 3:} RT-VNE chooses a group of X VNRs, embedding process is started from parent node (\textbf{Step 2}) and virtual nodes in each VNR are mapped sequentially in to other substrate nodes which are chosen by (Eq. 13).
		\STATE \textbf{Step 4:} Link Mapping is done by the shortest paths under the constraints (Eq. 13), selected paths must have enough bandwidth capacity as well as satisfy bandwidth capacity requirement ratio (Eq. 13).
		
	\end{algorithmic}
\end{algorithm}

\subsection{Virtual Node Mapping}	
In this step, we perform the node mapping process which consider how to map virtual nodes on substrate nodes. Because node mapping affects directly to link mapping,  we want to perform node mapping firstly. And then, link mapping will be successful and optimized. 

To achieved this goal, the key idea of our node mapping algorithm is that we first measure the CPU capacity and bandwidth resource of virtual nodes and substrate nodes and then we collect a group of \textit{X} VNRs and process $X$ VNRs sequentially in each scan-time window. 	

In term of virtual nodes, when a VNR arrives, we need to determine the mapping sequence of virtual nodes to the available substrate nodes. We define the \textit{VN-Node-Capacity} for virtual nodes $n^V$,  as follows:
\begin{equation}
Cap^{VN}(n^V) = C^V_{Req}(n^V) + \varPsi \sum_{l^V \in L^V(n)} BW^V_{Req}(l^V)
\end{equation} 
First of all, over each time interval, the topology of each VNR is detected by using BFS (Breath-First Search) algorithm, then virtual nodes are sorted in ascending order of hop distance from the parent virtual node which has highest required \textit{VN-Node-Capacity} value (Eq. 10). The parent virtual node is firstly embedded into substrate network due to the difficulty when we map the virtual nodes have high requirement CPU capacity and have many corresponding virtual links. For example, we use Figure. \ref{fig:figure3}(a) to illustrate for the process. That is, \textit{VN-Node-Capacity} value of virtual node \textit{a} is: $6 + 4 + 5 = 15$ unit, the value of virtual node $b$ and virtual node $c$ are 9 and 10, respectively. Virtual node $a$ has the highest \textit{VN-Node-Capacity} value, virtual node $a$ is parent virtual node and the embedding sequence is $a \to c \to b$.  Second, after choosing parent virtual node for each VNR, we determine a parent substrate node for each parent virtual node respectively. Because we selected $X$ parent virtual nodes of X VNRs so $X$ parent substrate nodes also should be selected. We use evaluation metric \textit{SN-Node-Capacity} value of parent substrate nodes  which is similar with parent virtual nodes. And the order of substrate nodes is sorted in decrease order of \textit{SN-Node-Capacity} values.
\begin{equation}
Cap^{SN}(n^S) = C^S_{RemA}(n^S) +	\varPsi\sum_{l^S \in L^S(n^S)}BW^S_{RemA}(l^S)
\end{equation}
where $L^S(n^S)$ is the set on adjacent substrate links of substrate node $n^S$.
In fact, there are many substrate nodes which have resource redundantly to become parent node. However, we prefer to select the parent node which have resource enough. As prior study\cite{6}, we also do not use $C(n^S)$ alone as an evaluation metric, because we consider to make sure that substrate node $n^S$ has enough CPU capacity available. Similarly, we use Figure. 3b to illustrate for the process, the \textit{SN-Node-Capacity} value of substrate node $3$ is highest ($10 + 5 + 7 + 9+ 6 = 37$), so that node $3$ is the parent substrate node for VNR which has the parent virtual node $a$.
\begin{figure}[t]
	\centering
	\includegraphics[width=0.5\linewidth]{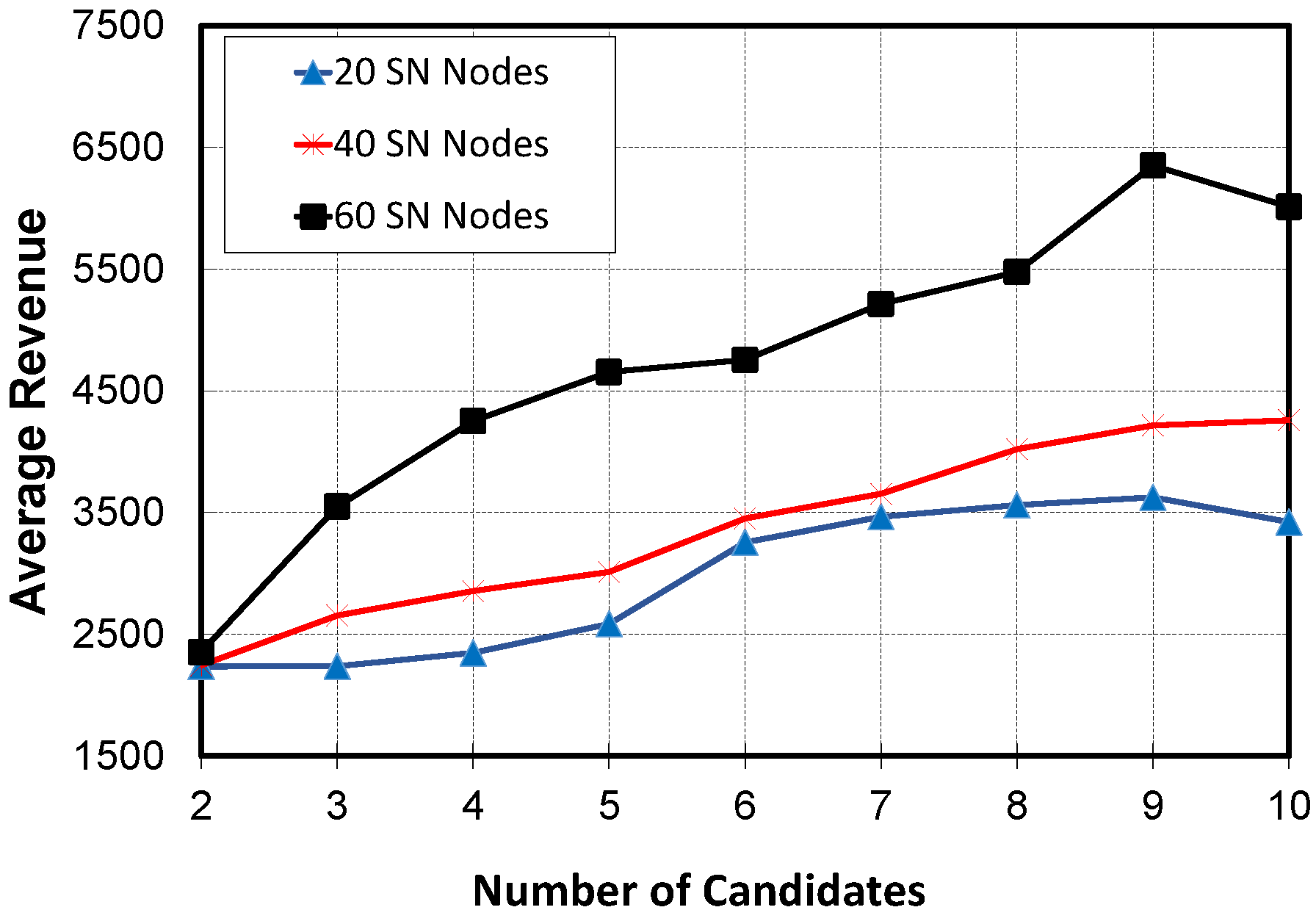}
	\caption{Impact of the number of candidates}
	\label{fig:figure4}
\end{figure}
\subsection{Virtual Node-Link Mapping Conjunction}	
Now we explain how to embed virtual nodes and links simultaneously onto substrate network. First of all, we define a metric called \textit{Link-traffic-ratio} $Ltr(l^S)$ which implies the ratio between a substrate link $l^S \subset L^S$ and a total bandwidth of all substrate links. The $Ltr(l^S)$ represents the intensity of overall interference that the SN link $l^S$ affects its other neighboring links. The \textit{Link-traffic-ratio} of a substrate path is the sum of the \textit{Link-traffic-ratio} of all links in the path. 
\begin{equation}
Ltr(l^S) = \frac{BW^S(l^S)}{\sum_{\forall l^S \in L^S} BW^S(l^S)}
\end{equation}
Figure. \ref{fig:figure3} shows an example of \textit{Link-traffic-ratio}. First of all, we assume that parent substrate node $n^S_i$ and parent virtual node $n^V_i$ of VNR $g^V_i \subset G^V$ as mentioned above. The rest of virtual nodes $n^V_j$ are embedded into corresponding substrate nodes and virtual links between virtual nodes $n^V_j$ are mapped following the shortest paths in terms of \textit{Link-traffic-ratio} according to each path has to satisfy bandwidth capacity requirements of virtual links. The node mapping rule is designed to minimize the load stress added by mapping a new virtual nodes to substrate nodes. In fact, we aim to choose the reasonable substrate nodes and shortest path offering the best bandwidths in order to reduce resource allocated to substrate links. Selecting substrate nodes $n^S_j$ for corresponding $n^V_j$ as follows: 

	\begin{equation}
	n^S_j \in argmin  \sum_{n^V_k \in N^V, n^S_k \in N^S} BW^S(n^V_k, n^V_j) Ltr(n^S_k,n^S_j).
	\end{equation}
	\begin{equation}
	\textrm{s.t. Bandwidth demands by added VN links. }
	\end{equation}

The equation (12) also shows the bandwidth capacity requirement ratio to map virtual link between a pair of virtual node $n^V_k$ and $n^V_j$.	For example, we use Figure. \ref{fig:figure3} to illustrate the process. In this example, we consider one VNR (Figure. 3a), where the parent virtual node $a$ and parent substrate node $3$. As mentioned, the next virtual node is node $c$, we need to find the corresponding substrate node which has the smallest \textit{Link-traffic-ratio} value from parent substrate node $C$ and satisfy bandwidth requirement of virtual link $(a,c)$, that is substrate node $5$, so virtual link $(a,c)$ is embedded to substrate link $(3,5)$. Finally, we find the available substrate node  to map virtual node $b$. Following the equation (12), we calculate the bandwidth capacity requirement ratio from substrate node $3$ and $5$ to other substrate nodes $1$, $2$ and $4$. In terms of substrate node $1$: $2*0.084+4*0.15=0.768$, for substrate node $2$: $2*(0.13+0.1)+4*0.13= 0.98$, for SN node $4$: $2*0.1+4*0.11=0.64$. So we select substrate node $4$ to map virtual node $b$, and virtual links $(b,c)$ and $(b,a)$ are embedded to substrate links $(4,5)$ and $(4,3)$, respectively. The process of our proposed algorithm called \textit{RT-VNE} Algorithm is described in Algorithm \ref{algorithm1}. In terms of VNR is embedded unsuccessfully, they then returned to queue and wait for next time interval (Figure. \ref{fig:figure2}).

\begin{figure}[t!]
	\centering
	\includegraphics[width=0.5\linewidth]{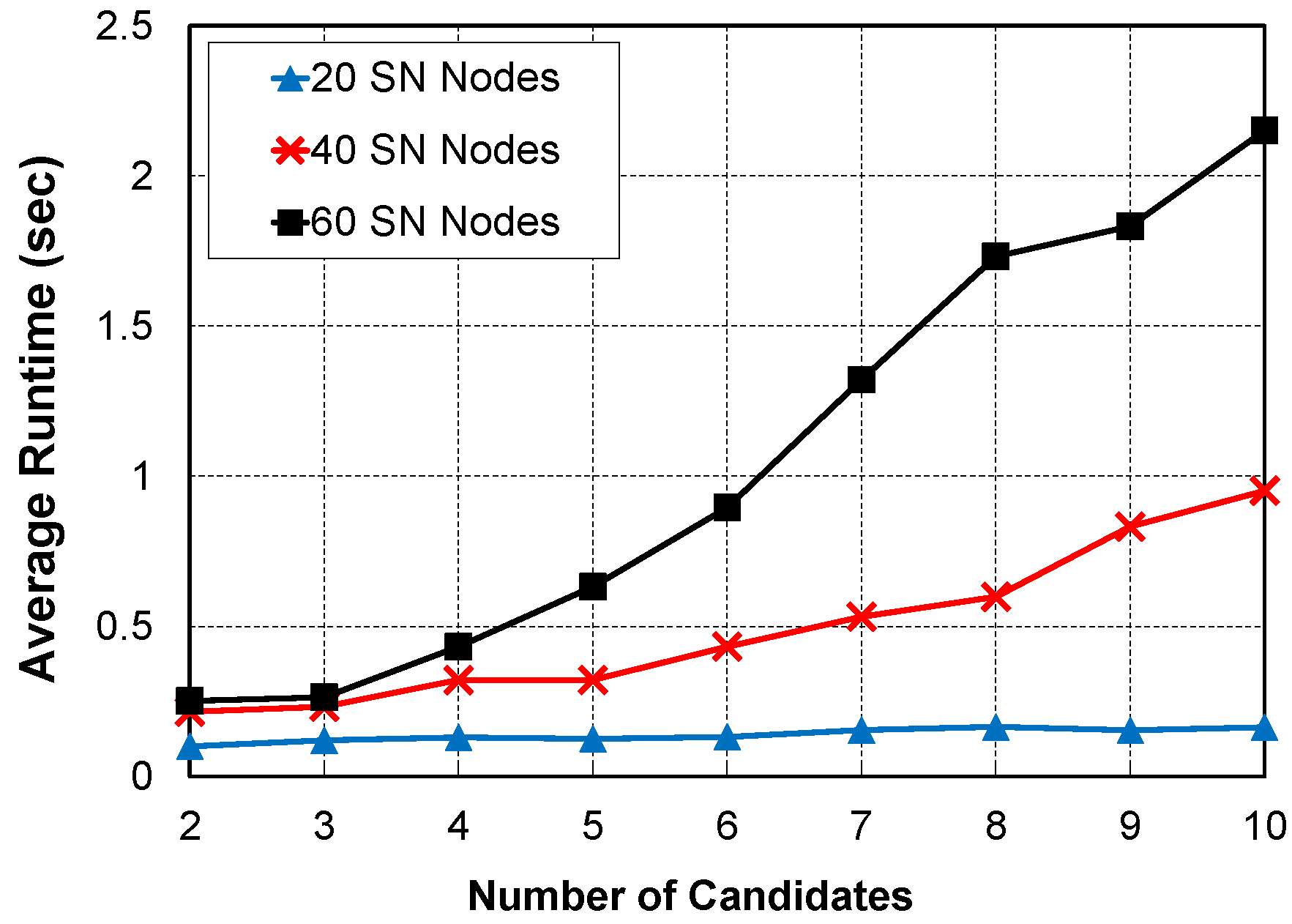}
	\caption{Average Run time per VN request}
	\label{fig:figure5}
\end{figure}

\section{Performance Evaluation}
\label{Evaluation}
In this section, we evaluate the performance of our proposed VNE algorithm in detail and compare to the prior algorithms under various scenarios. Our evaluation concentrates on quantifying the advantage of coordinate node and link mapping phases as well as proposed dynamic VNRs handling in terms of long-term average revenue and acceptance ratio. All simulations are performed on the server Intel(R) Core (TM) 2 Quad C Q9550 @ 2.83GHz (4Cs), 8GB memory, 1 TB disk and OS Linux 2.6.

\subsection{Simulation Environment} 
\textbf{Substrate Network:} Similar to most of prior studies, we used the GT-ITM tool \cite{26} to generate substrate network and virtual network topology. The substrate network is composed of 100 nodes, close to mid-size of an ISP scale. These substrate nodes are distributed geographically in a (10x10) grid square. Each pair of substrate nodes is connected with probability 0.5. The CPU capacity of nodes and bandwidth capacity of links in substrate network follow a uniform distribution from 0 to 300 unit.
\begin{table}[h]
	\small\sf\centering
	\caption{Compared Algorithm.\label{T2}}
	\begin{tabular}{p{1.7cm}  p{5.5cm}}
		\hline
		\textbf{Notation} & \textbf{Description}  \\ 
		\hline
		G-SP\cite{9} & Greedy Node Mapping with Shortest Path based Link Mapping \\	
		\hline					
		CC-VNE\cite{6} & Cluster Center VNE approach \\
		\hline			
		RT-VNE & Proposed Scheme \\				
		\hline
		R-ViNE\cite{11} & Randomized Node mapping with splittable Link Mapping using MCF\\	
		\hline
		D-ViNE\cite{11} & Deterministic Node Mapping with splittable Link Mapping using MSF \\		
		\hline
		D-ViNE-SP\cite{11} & Deterministic Node Mapping with Shortest Path based Link Mapping \\			
		\hline
	\end{tabular}
\end{table}		

\textbf{Virtual Network Request:} We evaluate the performance of VNE algorithms in various cases of VNRs. In each virtual network, the number of virtual nodes is distributed uniformly from 2 to 10 nodes for small size, from 10 to 50 nodes for mid-size and from 50 to 80 nodes for large size. The average of virtual network connectivity is set to be probability 0.5. Similar to substrate network configuration, the CPU capacity and bandwidth capacity of each virtual node and link are uniformly distributed between 0 and 30 unit. The arrival of VNRs is followed by Poisson process with an average which ranges from 2 to 14 VNRs per scan-time window. The lifetime of each VNR at the substrate network follows an exponential distribution with an average of 5 time windows. In each our simulation, there are approximately 500 scan-time windows, which correspond to around 2500 VNRs on average. The value of candidate VNRs X is varied from 2 to 10 (the default value is 9).  Figure. \ref{fig:figure4} shows the long-term average revenue as $X$ (number of searching candidates) increases from 2 to 10. We vary the number of substrate nodes with 20 nodes, 40 nodes and 60 nodes. Figure. \ref{fig:figure4} shows that with the value of X increases from 6 to 10 and the average revenue achieves good results. In Figure. \ref{fig:figure5}, the computation time also increases sharply with the increasing number of SN nodes, because the number of substrate links increases due to the fixed substrate network density. The value of searching candidate X might depend on the substrate network and virtual network sizes. The result implies that a small search number out of the large search space may be sufficient enough. Figure. \ref{fig:figure4} and Figure. \ref{fig:figure5} also demonstrate that our proposed algorithm is computationally tractable, because the proposed algorithm generates the mapping results in the order of seconds which is reasonable in real-world environment
\begin{figure}[t!]
	\centering
	\includegraphics[width=0.5\linewidth, height=0.3\textheight]{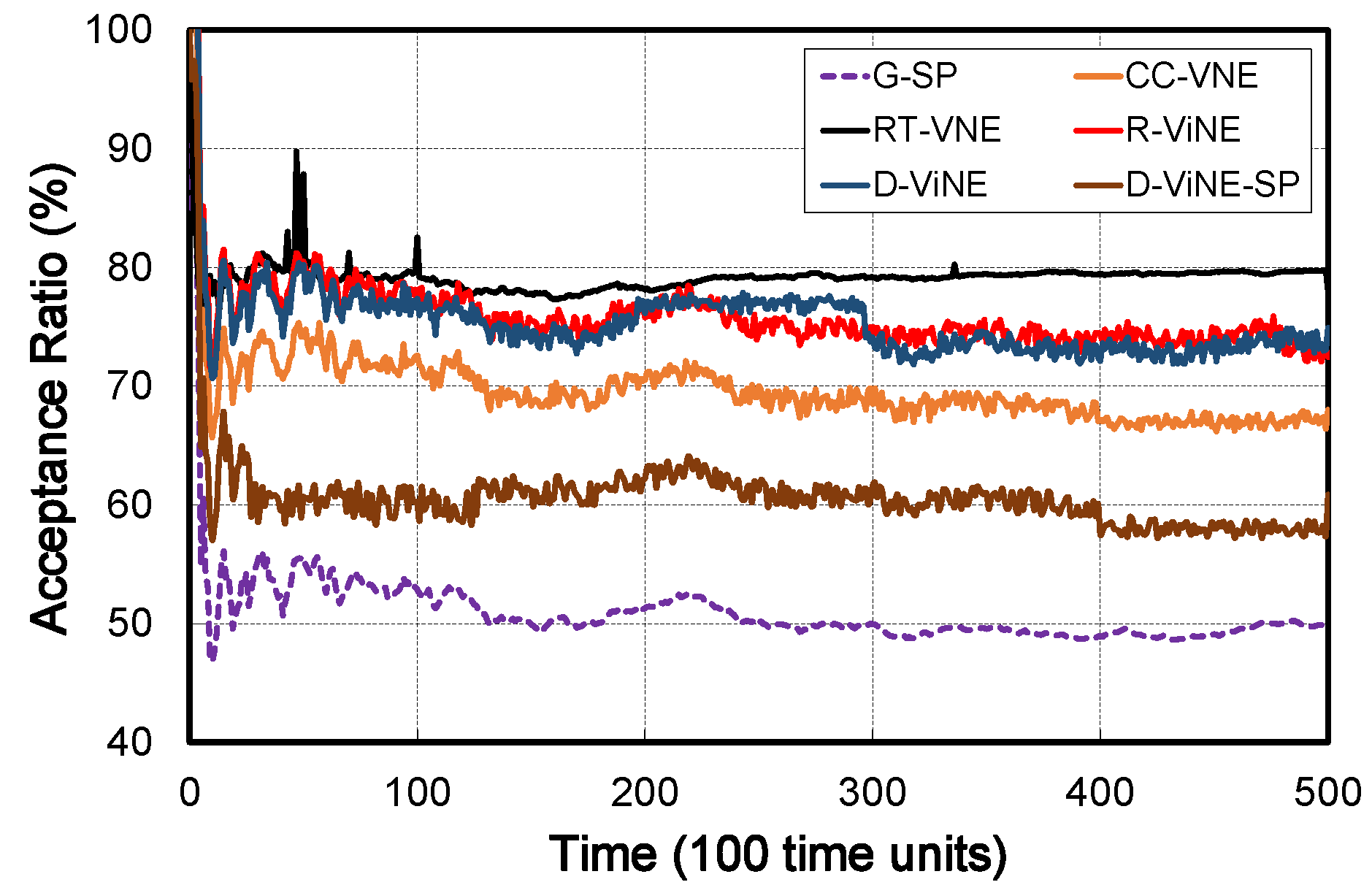}
	\caption{Acceptance Ratio of algorithms}
	\label{fig:figure6}
\end{figure}
\begin{figure}[h]
	\centering
	\includegraphics[width=0.5\linewidth]{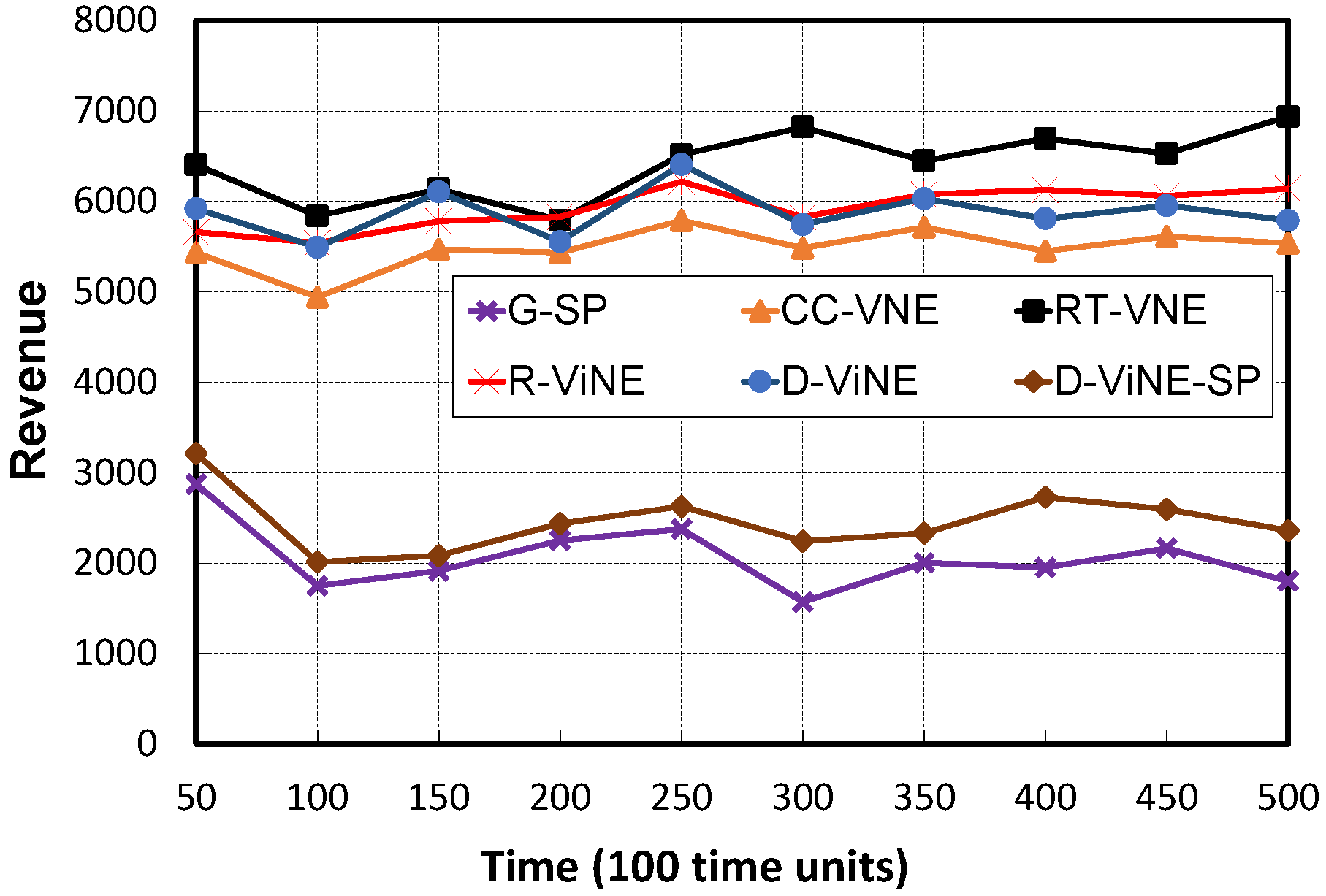}
	\caption{Average Revenue of algorithms}
	\label{fig:figure7}
\end{figure}
\begin{figure}[t]
	\centering
	\includegraphics[width=0.5\linewidth]{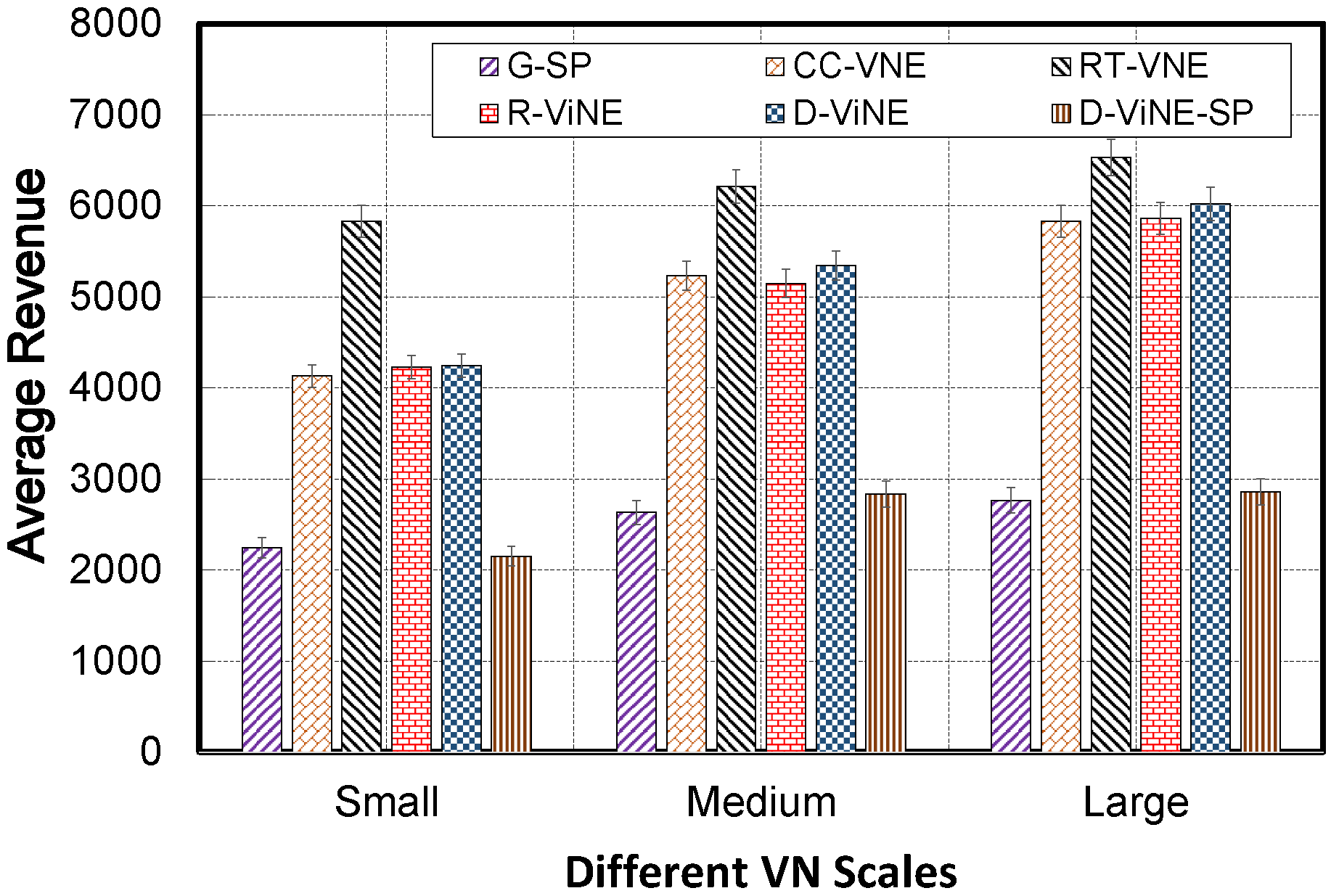}
	\caption{Average Revenue between algorithms on different VN scales}
	\label{fig:figure8}
\end{figure}

\textbf{Comparison Setup:} Our proposed algorithm is implemented in C/C++ and performed comparison to previous state-of-art algorithms. To evaluate the performance of our proposed algorithm RT-VNE, we compare our algorithm to existing algorithms (Table. \ref{T2}) included greedy mapping algorithm\cite{9}, a cluster-center VNE algorithm\cite{6}, and three ViNE algorithms\cite{11} . In greedy mapping algorithm (G-SP), virtual nodes with higher priority are mapped to substrate nodes with higher resource, and virtual links are embedded using\textit{ k-shortest-path} algorithm without reconfiguration. And the VNE algorithm CC-VNE\cite{6}, the authors first determine the cluster of substrate nodes which will be served virtual nodes. Then, they consider which substrate node has highest available resource is denoted as the cluster center and the other substrate nodes are sorted based on the distance from selected substrate nodes. Finally, the virtual nodes are mapped into selected substrate nodes such that higher degree virtual nodes are embedded to corresponding higher neighborhood resource availability substrate nodes.

\subsection{RT-VNE Algorithm Performance}

\textbf{Long-term Average Revenue:}  We compare the proposed VNE algorithm RT-VNE to five existing algorithms included G-SP, CC-VNE, R-ViNE, D-ViNE and D-ViNE-SP through 500 time windows. In our scenarios, VNRs are divided into three groups: small size [2,10] nodes, medium size [10,50] nodes and large size [50,80] nodes. From Figure. \ref{fig:figure6}, Figure. \ref{fig:figure7} and Figure. \ref{fig:figure8}, we have several interesting observations. Firstly, from the result in Figure. \ref{fig:figure7}, RT-VNE leads to better long-term average revenue than the existing algorithms in three cases small, medium and large size in the range of 500 scan-time windows. Along with these results, it implies that our proposed algorithm embeds VNRs that makes more benefit revenue without taking care the size of VNRs (small, medium or large) and increases the acceptance ratio of embedding. We believe the reason for it is that the higher acceptance ratio of RT-VNE (Figure. \ref{fig:figure6}) compared to existing algorithms as well as mapping virtual nodes to the substrate nodes with adequate resources increase the probability of satisfying the resource requirements of the VNRs and consequently helps to accelerate convergence. Moreover, in addition to node mapping constraints, RT-VNE also considers the link bandwidth constraint which obviously make link mapping phase easier. Thus, RT-VNE algorithm achieves higher probability of accepting VNRs. 

\textbf{Acceptance Ratio:} As shown in Figure. \ref{fig:figure6}, the proposed algorithm, RT-VNE leads to higher acceptance ratio compared to existing algorithms through joint node and link mapping. The acceptance ratio of RT-VNE (over 80\%) is stable and higher than others (G-SP is around 52\%, CC-VNE is about 72\% and R-ViNE, D-ViNE and D-ViNE-SP are around 76\%, 78\% and 63\%, respectively). Since the LP relaxation and rounding techniques used by \{R$|$D\}-ViNE algorithms may incur infeasible VNE solution, so they suffer low performance in term of VNR acceptance ratio, and consequently result in lower long-term average revenue(Figure. \ref{fig:figure7} and Figure. \ref{fig:figure8}). And G-SP and CC-VNE always select maximum substrate objects for embedding. Thanks to \textit{Link-traffic-ratio}, and the effect of combining node and link mapping is magnified when the available link resources are not enough. On the other hand, the influence of link interferences on embedding is not changed significantly since the relative among resources remain the same. And VN consolidation increases the efficiency of SN resource usage and save more available resources for other future VNRs. 
\begin{figure}[t!]
	\centering
	\includegraphics[width=0.5\linewidth]{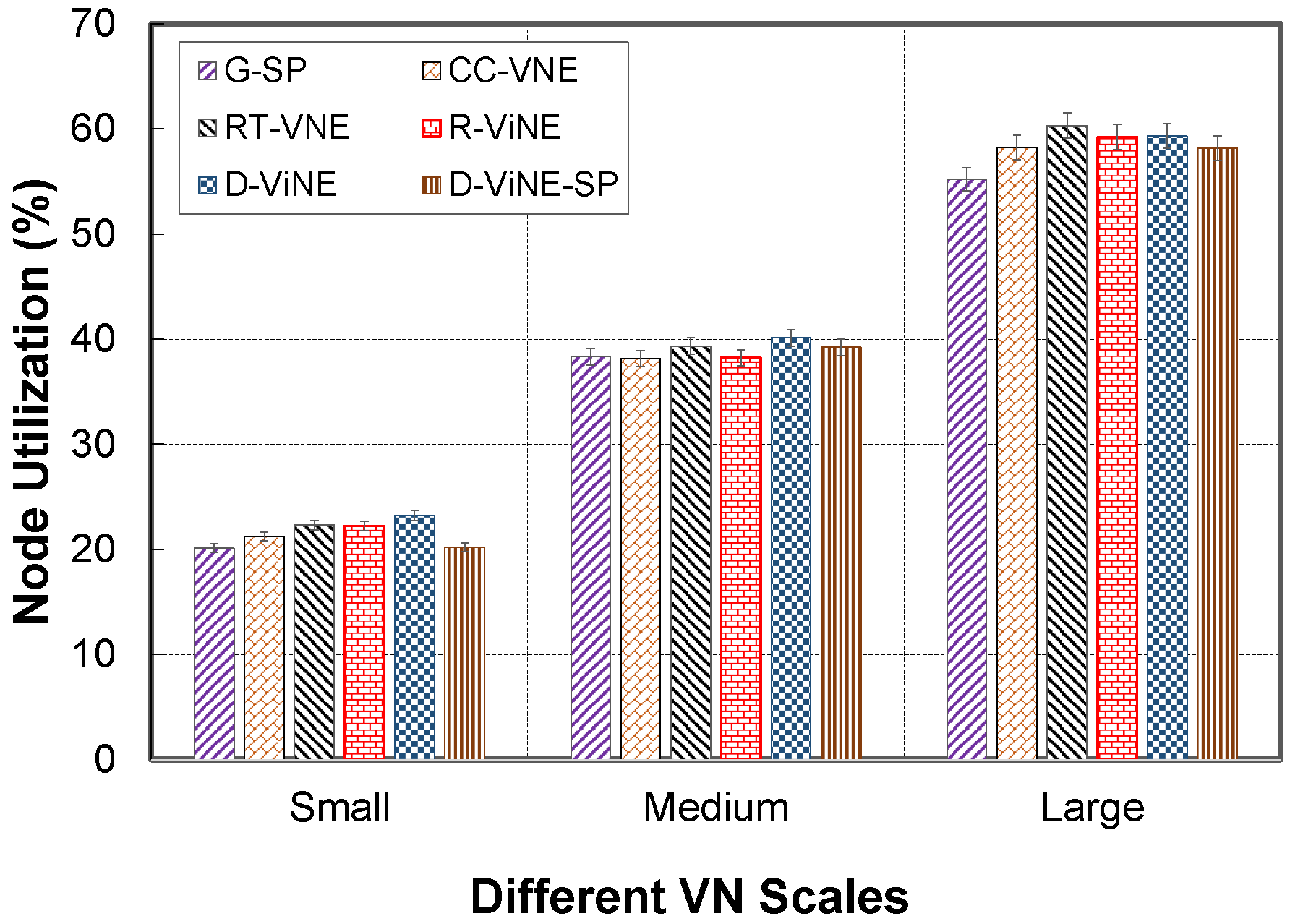}
	\caption{Average Node Utilization on different VN scales}
	\label{fig:figure9}
\end{figure}			

\begin{figure}[t!]
	\centering
	\includegraphics[width=0.5\linewidth]{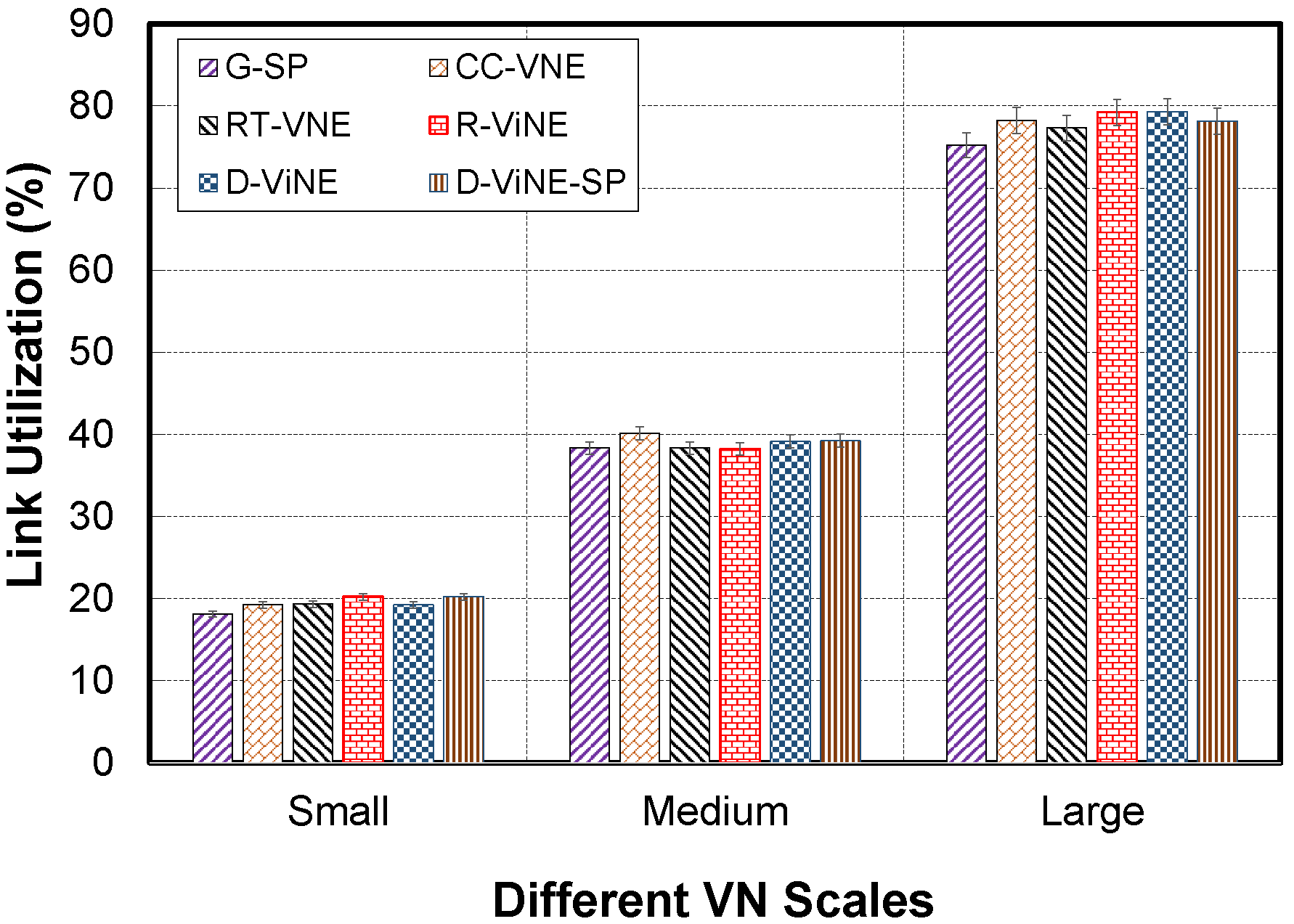}
	\caption{Average Link Utilization on different VN scales}
	\label{fig:figure10}
\end{figure}

\textbf{Resource Utilization:} The average resource utilization of substrate nodes and substrate links from different VNE algorithms are depicted in Figure. \ref{fig:figure9} and Figure. \ref{fig:figure10}. In terms of small, medium and large scales, the values of node utilization are around 20\%, 38\% and 58\%, respectively. The values of link utilization are approximately 20\%, 40\% and 79\%, respectively. Thanks to our \textit{Link-traffic-ratio}, even the acceptance ratio of RT-VNE is higher than others, the average node and link utilization of RT-VNE are quite the same with others. Consequently, network utilization is enhanced. As the increase of VN scale, all compared algorithms have higher node and link utilization. This is because there are more substrate nodes which have to active to handle incoming VN nodes as well as the number of SN links increases when VN scale increases. 

\textbf{Summary:} We can see that our proposed VNE RT-VNE achieves higher benefit revenue and acceptance ratio than state-of-art existing algorithms (Figure. 6-8). This can be explained by the fact that RT-VNE accepts many more VN requests than related approaches. Additionally, coordinated nodes and links mapping stages also lead to efficient mapping and consequently generates a higher income. We conclude that the effect of joint node and link mapping is significant. The proposed algorithm considers the neighborhood influence of substrate nodes, so with the limited embedding candidate number, it is highly choose efficient embedding candidates. Besides, the proposed algorithm checks the feasibility of embedding candidates before embedding, it is necessary for efficient embedding. 

\section{Conclusion}
\label{sec:conclusion}
Virtual Network Embedding is a critical key of Network Virtualization technology. This paper study the Virtual Network Embedding problem which is NP-hard and computationally intractable. In response, we propose a new resource efficient algorithm for VNE which coordinates node and link mapping phases and handles the online dynamic virtual network request in real-world, resulting in the improvement of embedding efficiency. The incoming VNRs are grouped, analyzed and ordered follow a selection rule. In each VNR, a VN node is selected as a root node and then mapped to root SN node into SN network. In comparison to some previous algorithms, the new strategies can markedly improve both resource utilization and long-term average revenue while the complexity is kept in an acceptable level.
%


\begin{thebibliography}{99}
	\bibliographystyle{SageV}
	
	\bibitem{1}
	Bhardwaj S, Jain L, Jain S. Cloud computing: A study of infrastructure as a service (IAAS). International Journal of engineering and information Technology. 2010 Feb;2(1):60-3.
	
	\bibitem{2}
	Fischer, Andreas, et al. "Virtual network embedding: A survey." IEEE Communications Surveys  Tutorials 15.4 (2013): 1888-1906.
	
	\bibitem{3}
	Chowdhury, N. M., and Raouf Boutaba. "Network Virtualization: state of the art and research challenges." Communications Magazine, IEEE 47.7 (2009): 20-26.
	
	\bibitem{4}
	Bianchi, Francesco, and Francesco Lo Presti. "A Markov Reward based Resource-Latency Aware Heuristic for the Virtual Network Embedding Problem." ACM SIGMETRICS Performance Evaluation Review 44, no. 4 (2017): 57-68.
	
	\bibitem{5}
	Ricci R, Alfeld C, Lepreau J. A solver for the network testbed mapping problem. ACM SIGCOMM Computer Communication Review, 2003, 33(2): 65−81.
	
	\bibitem{6}
	Zhu, Yong, and Mostafa H. Ammar. "Algorithms for Assigning Substrate Network Resources to Virtual Network Components." INFOCOM. Vol. 1200. No. 2006. 2006.
	
	\bibitem{7}
	Chowdhury, Mosharaf, Muntasir Raihan Rahman, and Raouf Boutaba. "Vineyard: Virtual network embedding algorithms with coordinated node and link mapping." IEEE/ACM Transactions on Networking (TON) 20.1 (2012): 206-219.
	
	\bibitem{8}
	D. Andersen, “Theoretical approaches to node assignment,” Unpublished manuscript, http://www.cs.cmu.edu/∼dga/papers/andersen-assign. ps, 2002
	
	\bibitem{9}
	Yu, M., Yi, Y., Rexford, J. and Chiang, M., 2008. Rethinking virtual network embedding: substrate support for path splitting and migration. ACM SIGCOMM Computer Communication Review, 38(2), pp.17-29.
	
	\bibitem{10}
	Zhang, Z., Su, S., Zhang, J., Shuang, K.,  Xu, P. (2015). Energy aware virtual network embedding with dynamic demands: Online and offline. Computer Networks, 93, 448-459.
	
	\bibitem{11}
	Chowdhury, N. M., Rahman, M. R.,  Boutaba, R. (2009, April). Virtual network embedding with coordinated node and link mapping. In INFOCOM 2009, IEEE (pp. 783-791). IEEE.
	
	\bibitem{12}
	Lischka, Jens, and Holger Karl. "A virtual network mapping algorithm based on subgraph isomorphism detection." In Proceedings of the 1st ACM workshop on Virtualized infrastructure systems and architectures, pp. 81-88. ACM, 2009.
	
	\bibitem{13}
	Cheng, Xiang, et al. "Virtual network embedding through topology-aware node ranking." ACM SIGCOMM Computer Communication Review 41.2 (2011): 38-47.
	
	\bibitem{14}
	Zhang, Sheng, et al. "Virtual network embedding with opportunistic resource sharing." IEEE Transactions on Parallel and Distributed Systems 25.3 (2014): 816-827.
	
	\bibitem{15}
	Su, Sen, et al. "Energy-aware virtual network embedding." IEEE/ACM Transactions on Networking 22.5 (2014): 1607-1620.
	
	\bibitem{16}
	Heller, Brandon, et al. "ElasticTree: Saving Energy in Data Center Networks." NSDI. Vol. 10. 2010.
	
	\bibitem{17}
	Beck, M.T., Fischer, A., de Meer, H., Botero, J.F. and Hesselbach, X., 2013, June. A distributed, parallel, and generic virtual network embedding framework. In Communications (ICC), 2013 IEEE International Conference on (pp. 3471-3475). IEEE.
	
	\bibitem{18}
	Beck, M.T., Fischer, A., Botero, J.F., Linnhoff-Popien, C. and de Meer, H., 2015. Distributed and scalable embedding of virtual networks. Journal of Network and Computer Applications, 56, pp.124-136.
	
	\bibitem{19}
	Fajjari, I., Saadi, N.A., Pujolle, G. and Zimmermann, H., 2011, June. VNE-AC: Virtual network embedding algorithm based on ant colony metaheuristic. In Communications (ICC), 2011 IEEE International Conference on (pp. 1-6). IEEE.
	
	\bibitem{20}
	J. Lu, J. Turner, Efficient mapping of virtual networks onto a shared substrate, 	Washington University in St. Louis, Tech. Rep (2006).
	
	\bibitem{21}
	Cheng X, Su S, Zhang Z, Shuang K, Yang F, Luo Y, Wang J. Virtual network embedding through topology awareness and optimization. Computer Networks. 2012 Apr 19;56(6):1797-813.
	
	\bibitem{22}
	Houidi, Ines, Wajdi Louati, and Djamal Zeghlache. "A distributed virtual network mapping algorithm." Communications, 2008. ICC'08. IEEE International Conference on. IEEE, 2008.	
	
	\bibitem{23}
	Houidi I, Louati W, Zeghlache D, Papadimitriou P, Mathy L. Adaptive virtual network provisioning. InProceedings of the second ACM SIGCOMM workshop on Virtualized infrastructure systems and architectures 2010 Sep 3 (pp. 41-48). ACM. 
	
	\bibitem{24}
	Guo T, Wang N, Moessner K, Tafazolli R. Shared backup network provision for virtual network embedding. InCommunications (ICC), 2011 IEEE International Conference on 2011 Jun 5 (pp. 1-5). IEEE.
	
	\bibitem{25}
	Duan, Jun, Zhiyang Guo, and Yuanyuan Yang. "Cost efficient and performance guaranteed virtual network embedding in multicast fat-tree DCNs." Computer Communications (INFOCOM), 2015 IEEE Conference on. IEEE, 2015.
	
	\bibitem{26}
	Zegura, E.W., Calvert, K.L. and Bhattacharjee, S., 1996, March. How to model an internetwork. In INFOCOM'96. Fifteenth Annual Joint Conference of the IEEE Computer Societies. Networking the Next Generation. Proceedings IEEE (Vol. 2, pp. 594-602). IEEE.
	
\end{thebibliography}
\end{document}